%% file: main.tex
\documentclass{vgtc}                          

\graphicspath{{figures/}{pictures/}{images/}{./}} 

\usepackage{times}                     

\usepackage{tabu}                      
\usepackage{booktabs}                  
\usepackage{lipsum}                    
\usepackage{mwe}                       

\usepackage{mathptmx}                  

\newcommand{\pheading}[1]{\vspace{4px}\noindent\textbf{#1}}

\newcommand{\toolkit}{\textsc{R\'{e}citKit}}

\onlineid{1008}

\vgtccategory{Research}

\vgtcinsertpkg

\title{\toolkit: A Spatial Toolkit for Designing and Evaluating\\ Human-Centered Immersive Data Narratives}

\author{
   Vidya Setlur\thanks{e-mail: vsetlur@tableau.com}\\ %
    \scriptsize Tableau, Palo Alto CA USA %
\and Samuel Ridet \thanks{e-mail: sridet@salesforce.com}\\ %
     \scriptsize Tableau, Palo Alto CA USA}

\teaser{
 \centering
 \includegraphics[width=.8\linewidth]{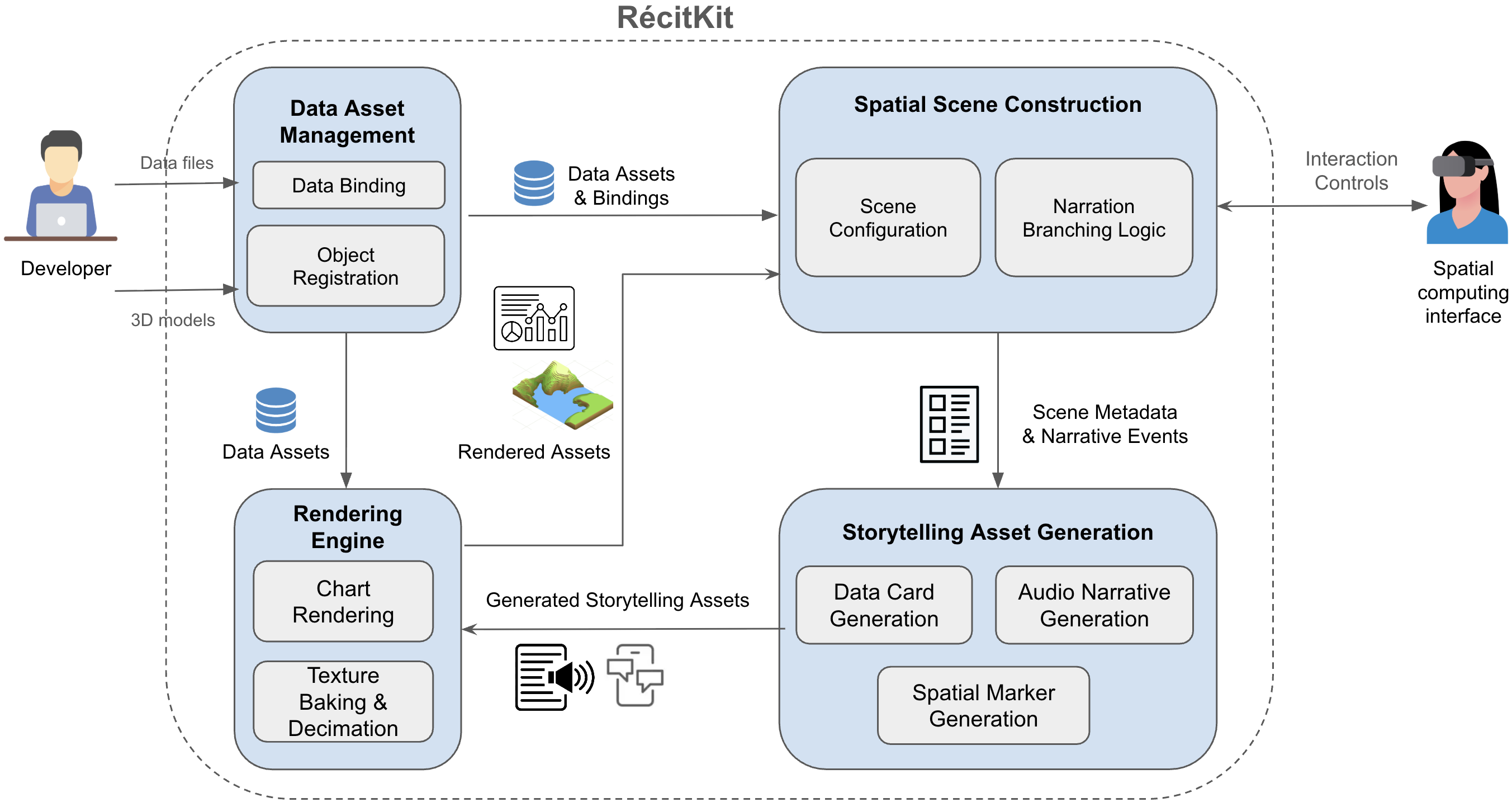}
 \caption{\toolkit~overview. The app developer supplies structured data files and 3D models as inputs to the toolkit. The \textit{Data Asset Management }module binds and registers assets, passing them to the S\textit{patial Scene Construction} module, which configures scenes and branching narratives. \textit{Storytelling Asset Generation} produces dynamic content such as data cards, spatial markers, and audio narratives, informed by scene metadata. The \textit{Rendering Engine} generates 3D models, immersive environments and narrative assets with optimizations like texture baking and decimation. Finally, interaction controls enable a user to explore and interact with the spatial data storytelling experience through a head-mounted display (HMD) interface.}
 \label{fig:teaser}
}

\abstract{
  \input{sections/00_abstract}
} 

\keywords{Immersive analytics, data storytelling.}

\begin{document}


\firstsection{Introduction}

\maketitle

\input{sections/01_introduction}

\section{Related Work}
\input{sections/02_relatedwork}

\section{\toolkit~Toolkit}
\input{sections/03_spatial_toolkit}

\section{Use Case Application for \toolkit}
\input{sections/04_minard_application}

\section{Preliminary Evaluation}
\input{sections/05_evaluation}

\section{Discussion and Future Implications}
\input{sections/06_discussion}

\clearpage
\bibliographystyle{abbrv-doi}

\bibliography{main}
\end{document}

%% file: sections/00_abstract.tex
Spatial computing presents new opportunities for immersive data storytelling, yet there is limited guidance on how to build such experiences or adapt traditional narrative visualizations to this medium. We introduce a toolkit, \toolkit~for supporting spatial data narratives in head-mounted display (HMD) environments. The toolkit allows developers to create interactive dashboards, tag data attributes as spatial assets to 3D models and immersive scenes, generate text and audio narratives, enabling dynamic filtering, and hierarchical drill-down data discoverability. To demonstrate the utility of the toolkit, we developed Charles Minard’s historical flow map of Napoleon's 1812 campaign in Russia as an immersive experience on Apple Vision Pro. We conducted a preliminary evaluation with 21 participants that comprised two groups: \textit{developers}, who evaluated the toolkit by authoring spatial stories and \textit{consumers}, who provided feedback on the Minard app's narrative clarity, interaction design, and engagement. Feedback highlighted how spatial interactions and guided narration enhanced insight formation, with participants emphasizing the benefits of physical manipulation (e.g., gaze, pinch, navigation) for understanding temporal and geographic data. Participants also identified opportunities for future enhancement, including improved interaction affordance visibility, customizable storytelling logic, and integration of contextual assets to support user orientation. These findings contribute to the broader discourse on toolkit-driven approaches to immersive data storytelling across domains such as education, decision support, and exploratory analytics.

%% file: sections/01_introduction.tex
The evolution of spatial computing and HMDs has opened new paradigms in digital storytelling, enabling data narratives to move beyond static visualizations into fully interactive, immersive experiences~\cite{visionpro,vive,metaquest,hololens}. Traditional data storytelling often relies on 2D visualizations and dashboards, which, while effective, can limit users' ability to fully grasp the multidimensional complexity of datasets~\cite{ens2021grand,zhou:2023}. Immersive data storytelling integrates principles from narrative visualization, virtual reality (VR), and augmented reality (AR) to create interactive experiences that blend exploration, interactivity, and structured storytelling ~\cite{sega1993megavisor,cave,vrbook,jain:2024}. Research in immersive analytics highlights the benefits of such environments in fostering deeper engagement, improving spatial understanding, and facilitating collaborative analysis~\cite{Isenberg2018ImmersiveVD}. However, designing effective immersive storytelling experiences requires new methodologies that incorporate principles from both traditional visual analytics and emerging immersive technologies~\cite{saffo2023unraveling,sicat:2019}. 

To address this gap, we introduce a spatial computing toolkit, \toolkit\footnote{Derived from the French word r\'{e}cit meaning ``narrative," the name reflects our goal of supporting data-driven story discovery.} designed to support the development of immersive data storytelling experiences. The toolkit facilitates immersive narrative construction by providing capabilities for loading data files, dynamically binding data to high-fidelity 3D models, and managing scene elements with geographic and spatial context. \toolkit~enables developers to link content through gesture-based interactions available on the spatial computing device. Beyond interaction, the toolkit offers mechanisms for generating dynamic data cards, i.e., contextual overlays that display detailed metadata or historical information linked to specific spatial markers, and supports AI-generated spatial audio narratives, aligning narration to the user's location and exploration sequence within the environment. Developers can tag dimensional columns in the underlying dataset as spatial assets or real-world objects, supporting dynamic filtering, hierarchical grouping, and data discoverability without the need to ingest or manually interpret complex data structures. Additionally, \toolkit~simplifies the placement of spatial markers, which act as interactive waypoints to reveal context-sensitive information or trigger narrative transitions during exploration.

To demonstrate the capabilities of \toolkit, we developed a use case application: an immersive reinterpretation of Minard’s visualization of Napoleon’s March~\cite{minard1869} on the Vision Pro platform. This prototype served as a testbed to explore cognitive and physical interaction factors, such as spatial attention, comprehension, and recall within a structured yet explorable data narrative. We conducted a preliminary evaluation with 21 participants (app developers and consumers) to assess both the capabilities provided by \toolkit, such as interactive spatial navigation, dynamic storytelling, and multimodal data interaction, and their instantiation within the Minard use case application. The study surfaced actionable insights into human interaction patterns, highlighting both the strengths of immersive storytelling tools and key limitations around interaction affordances, scene transitions, and user orientation.

%% file: sections/02_relatedwork.tex
The field of immersive analytics and data storytelling, leveraging the capabilities of HMDs, has gained attention in recent years~\cite{ens2021grand,saffo2023unraveling}, with applications spanning domains including education~\cite{billinghurst:2012}, healthcare~\cite{pottle:2019}, and training~\cite{Webel2013AnAR}, to name a few. Chandler et al.~\cite{Chandler2015ImmersiveA} highlight the potential of immersive analytics to provide new ways of interacting with data, emphasizing the advantages of spatial data representation over conventional 2D visualizations. Similarly, Marriott et al.~\cite{marriott2018immersive} discuss how immersive environments can support complex data analysis tasks by enabling people to interact with data in more natural ways, such as through direct manipulation and spatial navigation. Building on this prior research, our work contributes a toolkit-based approach that operationalizes immersive storytelling principles within a spatial computing environment, supporting dynamic narrative adaptation, multimodal user interaction, and scene-aware data exploration, enabling app developers to author, structure, and deploy immersive data narratives.

Bowman et al.~\cite{bowmanbook} provide a framework for 3D user interface design, emphasizing the need for natural and intuitive interactions in AR/VR environments. These principles are particularly relevant for spatial storytelling, where users must fluidly transition between narrative sequences, exploratory analysis, and interactive data manipulation. Empirical research demonstrates that these input modalities can improve interaction precision and reduce cognitive load compared to traditional devices like a mouse and keyboard~\cite{Pfeuffer:2020,pfeuffer2024design}. Our work builds on these foundations by integrating gesture, gaze, and voice interaction into a unified toolkit for immersive data storytelling, enabling users to engage with spatial narratives. In addition to supporting natural interaction, \toolkit~introduces narrative-aware input handling, where user gestures or gaze selections dynamically influence storytelling flow, branching logic, and detail-on-demand retrieval of data facts and narrative content.

Prior research in data storytelling has established the foundations for integrating narrative structures with data visualization to enhance comprehension, engagement, and retention~\cite{segelheer2010,Isenberg2018ImmersiveVD}. More recently, immersive data-driven storytelling has emerged as an interdisciplinary field that blends immersive analytics with narrative techniques~\cite{jain:2024}. Zhou et al. highlight how hybrid storytelling approaches leverage both traditional 2D narratives and immersive 3D experiences to provide context-rich and interactive storytelling environments~\cite{zhou:2023}. In the context of immersive analytics, Sicat et al. present a toolkit designed for building immersive data visualizations, demonstrating how AR and VR technologies can be employed for data-driven presentations and storytelling~\cite{sicat:2019}. Mendez et al. explore how immersive storytelling can benefit from principles derived from game design, journalism, and education, proposing frameworks for structuring data stories in spatial computing environments~\cite{mendez}. While prior toolkits primarily focused on visualization construction or presentation models, our approach integrates spatial narrative elements—such as data-linked assets, audio narration, and interactive story branching into a cohesive experience optimized for head-mounted display environments.

Recent work has explored the intersection of immersive environments and data storytelling. Data visceralization proposes a method to enhance intuitive understanding of physical quantities through virtual reality environments by aligning data representations with real-world spatial mappings~\cite{newpaperlee:2020}. Techniques such as strollytelling couple physical locomotion with animated narrative progression to foster active engagement with immersive data stories~\cite{newpaper:jain2025}. Other work has examined the creative workflows and challenges associated with designing animated VR stories~\cite{newpaper:Yuan_2025}, as well as the effects of viewpoint (ego- vs. exocentric) and navigation (active vs. passive) modes on users' spatial immersion and understanding~\cite{newpaper:Lu_2025}. A recent paper surveys the emerging field of immersive data-driven storytelling, identifies key dimensions across research, journalism, and games, and calling for more structured frameworks to guide future development~\cite{newpaper:mendez2025}. Our work builds upon these important insights into immersive storytelling experiences by introducing a generalizable spatial computing toolkit, \toolkit~for authoring immersive, interactive data narratives. Rather than focusing on specific interaction techniques or narrative workflows, \toolkit~provides modular capabilities for loading structured datasets, binding data to 3D spatial markers, generating spatially-aware audio narratives, and dynamically sequencing storytelling scenes. We demonstrate the toolkit's utility by exploring how spatial computing can enhance comprehension, narrative recall, and user engagement when interacting with historical and analytical datasets.

%% file: sections/03_spatial_toolkit.tex
The motivation for the spatial computing toolkit stems from the need to support immersive data storytelling, drawing inspiration from prior research on narrative visualization, immersive analytics, and embodied interaction~\cite{Isenberg2018ImmersiveVD,sicat:2019, vanFossen:2023}. While general interface guidelines for visionOS~\cite{visionosguidelines} exist for rendering content legibly and responsively, such as maintaining appropriate text size, contrast, and interaction affordances, our work builds upon these insights by designing a spatial computing toolkit, \toolkit~that integrates best practices from visual analytics~\cite{few,mackinlayshowme} as well interactive storytelling techniques in immersive environments~\cite{saffo2023unraveling, marriott2018immersive}.

\toolkit~is implemented using the visionOS SDK~\cite{visionossdk}, incorporating SwiftUI, RealityKit, and ARKit for  2D and 3D rendering, gesture recognition, and gaze tracking. Figure~\ref{fig:teaser} illustrates the key components of \toolkit. An app developer first loads data files, corresponding storytelling assets, and 3D models into the \textit{Data Asset Management} module of the toolkit. This module updates a SQLite database~\cite{sqlite2020hipp} for structured data storage, organizing data attributes such as location, time, and event metadata into queryable schemas. Each data entry is bound to its corresponding visual asset (e.g., the 3D model) via a data binding layer that establishes persistent linkages between database fields and scene objects. This binding enables automatic real-time updates when users interact with data-driven elements in the HMD environment. Additionally, during asset import, objects are registered with the scene graph through an internal object registry system. This registry supports spatial indexing and semantic tagging of 3D objects, allowing for efficient runtime retrieval, interaction, and event triggering based on user gaze or gesture inputs, a technique commonly employed in AR/VR applications~\cite{arett:2021}. More concretely, data-bound assets are automatically augmented with unique IDs, position and orientation metadata, and interaction affordances. The \textit{Spatial Scene Construction} module orchestrates the dynamic configuration of immersive scenes and the branching logic of narrative events within the spatial storytelling environment. At runtime, this module parses the registered 3D objects and data assets to construct a coherent and semantically meaningful spatial layout. Each object or asset is associated with metadata such as spatial coordinates, orientation, and temporal sequencing obtained from the Data Asset Management module. Here is a format of the spatial scene construction graph generated by the \toolkit. Here, each scene object encodes an identifier, type, spatial position, attributes, and interaction handlers that map user actions to narrative events and scene transitions.

{\scriptsize
\begin{verbatim}
{
  "scene_id": "scene_identifier",
  "objects": [
    {
      "object_id": "unique_object_id",
      "type": "object_type",
      "position": { "x": value,
                    "y": value,
                    "z": value},
      "attributes": {
        "property_1": "value",
        ...
      },
      "interactions": {
        "on_gaze": "action_identifier",
        "on_gesture_select": "action_identifier",
        ...
      }
    }
  ],
  "narrative_sequence": [
    {
      "event_id": "event_identifier",
      "trigger": "trigger_type",
      "action": "action_to_perform",
      "next_scene": "optional_next_scene_id"
    }
  ]
}
\end{verbatim}}

The \textit{Storytelling Asset Generation} module produces immersive storytelling elements such as data cards, audio narratives, and spatial markers, based on the metadata and spatial scene construction graph. \toolkit~converts textual narratives from the data files into audio narratives using OpenAI's Text-to-Speech API~\cite{openai-tts}, where the developer can customize the voice characteristics via an input prompt in the toolkit. Scene configuration is driven by scene metadata and narrative event triggers attached to data elements. When a user interacts with a spatial marker (e.g., via gaze, gesture), the module updates the scene context by revealing new assets, repositioning entities, or triggering animations based on the branching structure defined in the narrative logic. Scenes are organized through the logical grouping of assets, supporting scalable spatial storytelling where localized details can be progressively disclosed within broader narrative arcs~\cite{SHNEIDERMAN2003364}. Transitions between narrative stages are managed through a state machine that encodes scene states, transition conditions, and animation sequences. Developers can specify branching points in \toolkit~where users' interactions dynamically alter the exploration of the underlying data story.

Finally, the \textit{Rendering Engine} processes these assets and visual models to generate 2D charts and dashboards, along with rendering 3D models and immersive scenes. The module applies decimation modifiers and texture baking, techniques that enhance rendering efficiency by reducing computational load while preserving visual fidelity~\cite{blenderdecimatemodifiers, texturebaking}. Decimation reduces the polygon count of 3D models, allowing for dynamic level-of-detail adjustments, which is particularly useful for rendering models in immersive environments. Texture baking precomputes lighting, shadows, and material properties onto a 2D texture map, eliminating the need for real-time lighting calculations and further improving performance.

%% file: sections/04_minard_application.tex
\begin{figure}[ht]
 \centering
  \includegraphics[width=\linewidth]{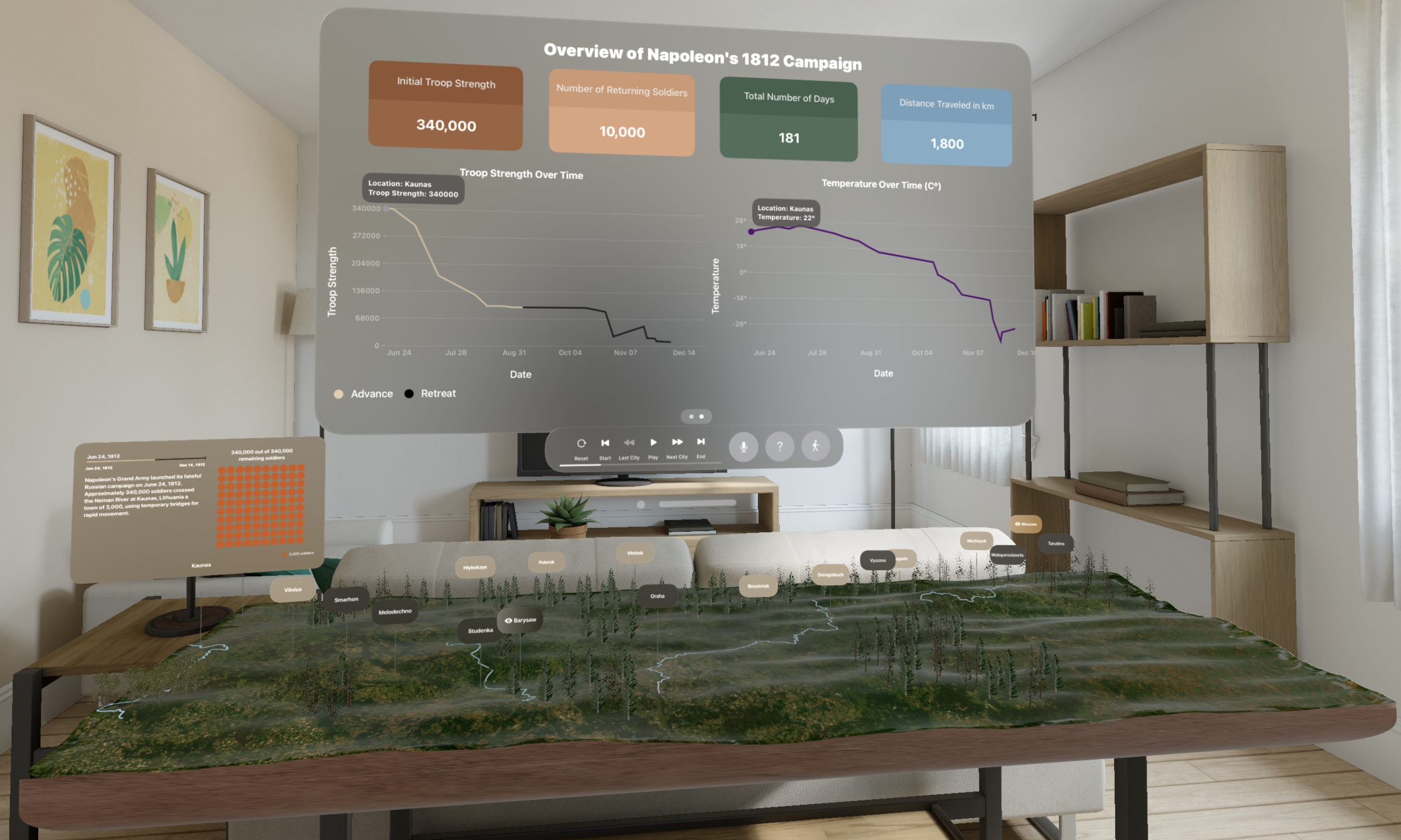}
 \caption{Minard application developed with \toolkit, integrates interactive 3D models, a dashboard, and gesture-based interactions to explore metrics such as troop strength, temperature variations, and campaign progress over time.}
 \label{fig:minard}
\end{figure}

To demonstrate the capabilities of the \toolkit, we employ the toolkit for developing a well-known historical data narrative, Minard’s visualization of Napoleon’s March to Moscow~\cite{minard1869}. 

The application provides two complementary perspectives: a dashboard view and a 3D terrain model, each mapping to specific modules within the \toolkit~architecture, as seen in Figure~\ref{fig:minard}. The dashboard presents an aggregated overview of the data comprising troop size, temperature, and historical milestones~\cite{minarddataset}. The 3D terrain model provides a detailed, spatially grounded view of the historical narrative. The Data Asset Management module supplies structured data bindings to the visual elements, while the Spatial Scene Construction and Rendering Engine modules instantiate troop movement paths, terrain features, and data overlays. Narrative sequencing and contextual annotations within the terrain are orchestrated through the Storytelling Asset Generation and Storytelling Engine modules, providing users with location-specific insights as they progress through the narrative experience.

\begin{figure}[ht]
 \centering
  \includegraphics[width=\linewidth]{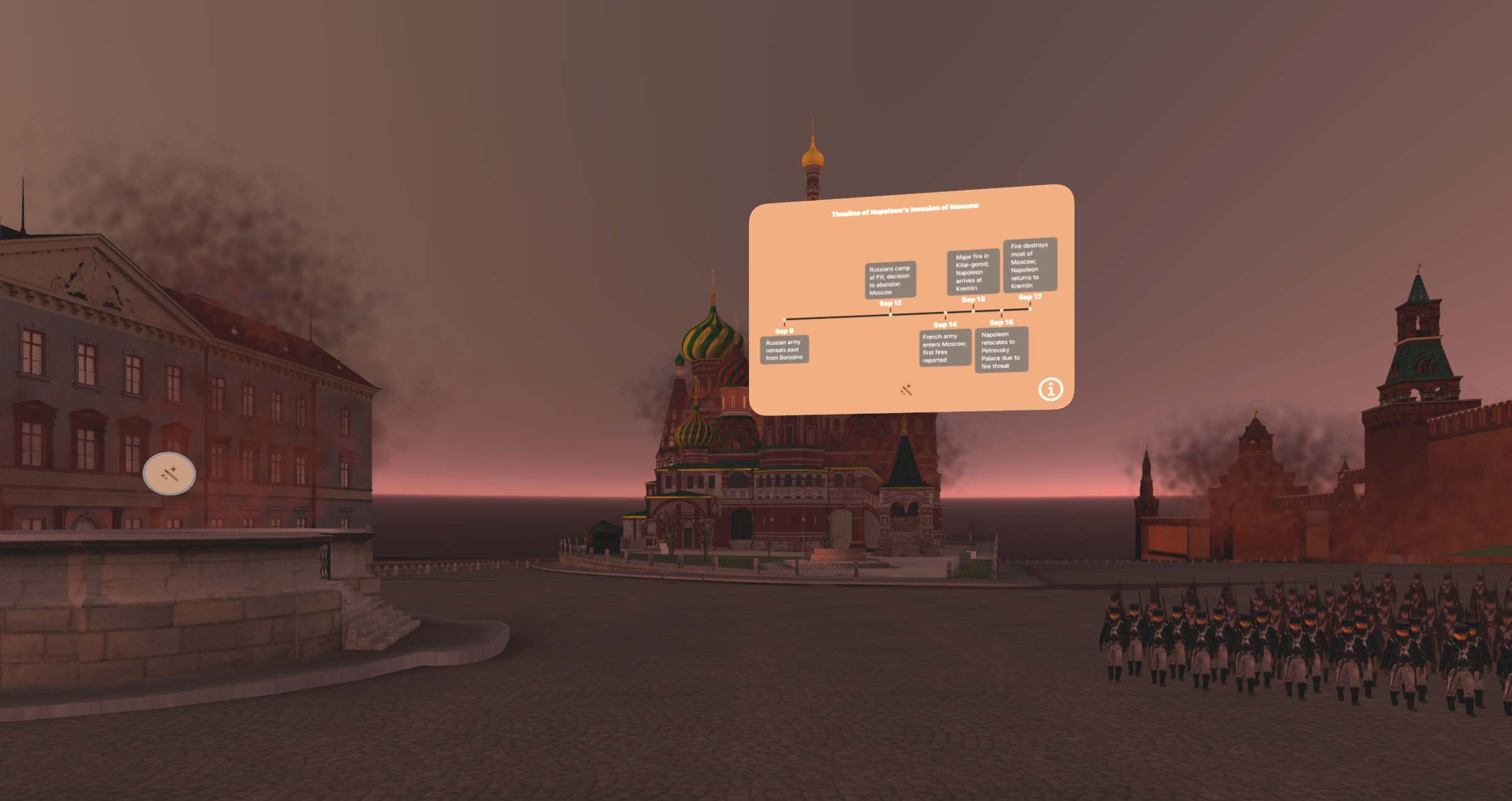}
 \caption{This interactive immersive view recreates Napoleon’s invasion of Moscow, depicting the city in flames after the Russians strategically abandoned the capital. Users can engage with spatially aware narratives by selecting clickable circular data markers, which reveal localized data card overlays beyond the background audio narration.}
 \label{fig:interface}
\end{figure}

To accommodate both structured storytelling and open-ended exploration, we leverage \toolkit~to support the following modes: (1) An auto-narrative playback that leverages the toolkit's support for scene sequencing and branching logic to guide users through a chronological journey of Napoleon’s March. (2) The exploration mode enables users to freely navigate the 3D terrain visualization, using gesture and voice controls to interact with various narrative elements. 
Users can then transition into localized, fully immersive scenes from the terrain model and interact with 3D spatial markers at key historical sites, such as St. Basil’s Cathedral in Moscow (Figure~\ref{fig:interface}).  A video showing a detailed walkthrough of the use case application is provided as supplementary material.

%% file: sections/05_evaluation.tex
To assess the effectiveness of \toolkit~for immersive data storytelling, we conducted a preliminary user study with a total of 21 participants. Participants comprised two groups: 9 self-reported as HMD application developers familiar with visionOS or comparable platforms ($D1 - D9$), and 12 were general consumers without development experience ($C10 - C21$), and their roles included data analysts, managers, solutions engineers, and market research consultants. Most participants were highly advanced in data analytics (12 participants), while others reported intermediate expertise (8 participants) or novice status (1 participant). Each study session lasted between 45 and 60 minutes. All participants were guided through device fitting and calibration to personalize eye tracking and hand gesture recognition on the Vision Pro. 

For the developer group, we asked participants to work with a global CO\textsubscript{2} emissions dataset~\cite{emissions} and construct a simple, immersive data story using a combination of chart components. Participants were encouraged to structure a narrative that guided users through temporal trends and geographic comparisons. Tasks included identifying the country with the highest CO\textsubscript{2} emissions in Asia, narrating the shift in global emission rankings over time, and highlighting the moment when South Korea’s emissions surpassed 500 million metric tons. These tasks allowed us to assess how developers used \toolkit's modules, such as data binding, storytelling asset generation, and spatial scene construction to sequence narrative events, apply data-driven filters, and configure immersive elements like spatial markers and annotations. Participants were provided with a laptop preloaded with the toolkit and dataset.

The consumers explored dynamic narratives, interpreted spatially embedded data, and engaged with immersive storytelling elements to support their analytical sense-making processes. They were asked to interact with the Minard application, completing tasks such as, manipulating the 3D terrain model, navigating structured storytelling sequences and accessing spatially embedded data details (i.e., troop movements, temperature changes, city-specific statistics). Finally, they were asked to explore the fully immersive experiences. Both sets of participants concluded the sessions by reflecting on either the toolkit or the application's user experience.

%% file: sections/06_discussion.tex
The user study provided valuable insights from both developers and consumers, offering perspectives on toolkit usability, storytelling engagement, and the broader potential for immersive data storytelling in spatial computing environments. We organize the feedback into key thematic areas that reflect participants' experiences with authoring spatial narratives using \toolkit~as well as interacting with the Minard application.

\pheading{Need for Data Storytelling Templates.} Developers found the configuration of \toolkit~components to be well-structured, especially the modular nature of data binding and scene construction. However, some participants noted that it still required familiarity with the scene objects--``\textit{It’s not hard to set up, but you need to understand the object model to get the bindings right.} [$D3$]" Participants echoed the value of providing template-based guidance in the future for rapidly assembling basic spatial data storytelling flows without having to build each component from scratch.

\pheading{Facilitating Data-driven Narrative Design.} Developers valued the toolkit’s support for authoring a wide range of data storytelling formats, including linear chronologies, spatially anchored scenes, and event-driven narratives. The ability to link scene transitions and narrative events to both spatial properties and temporal data fields enabled participants to construct stories based on either user interaction or the underlying dataset. One enterprise developer remarked, ``\textit{It’s easy to imagine adapting this for business scenarios; like walking through a sales pipeline or customer journey spatially, with data triggering what the story shows next.} [$D7$]"

\pheading{Scene Graph and Data Binding Previews.} Several participants appreciated the clarity of the spatial scene graph and the ability to directly map data fields to 3D object properties. However, they requested richer documentation and debugging tools to understand how data attributes influence runtime behavior. ``\textit{Seeing a live preview of how data drives changes in the environment would be a game-changer,}" noted $D4$, pointing to the potential of adding a real-time editor for linking data with scene actions.

\noindent We now present key themes that emerged from consumer participants' interactions with the Minard application:

\pheading{Immersive Contextualization Enhances Understanding.}
Participants noted that viewing historical data in a spatial context deepened their understanding of troop movements, temperature impact, and geographic scale. They found the terrain model and 3D troop paths to be particularly effective in connecting abstract data to physical locations. One participant shared, ``\textit{I’ve seen this chart before, but being able to follow the path across the map made it all click. I now understand the scale of the losses.} [$C13$]" Others emphasized that the immersive view helped them grasp the environmental conditions and narrative flow more intuitively than a static 2D version.

\pheading{Balancing Guidance and Exploration.}
While participants appreciated the guided auto-narrative mode, many also valued the freedom to explore story elements at their own pace. However, they emphasized that clear narrative anchors (such as event markers and summaries) were essential to avoid getting lost. ``\textit{I liked the ability to look around, but I still wanted the story to pull me back at key data points} [$C17$]." This suggests that spatial data stories benefit from a hybrid model that blends open-ended interaction with structured checkpoints.

\pheading{Visual and Structural Cues in Data Storytelling.}
Users highlighted the need for visual affordances to clarify narrative transitions, progression, and focal points. While the immersive visuals were compelling, participants often requested clearer indicators of where the story was going next. ``\textit{Sometimes I wasn’t sure if I’d reached the end of something or missed a point—there should be a better breadcrumb trail} [$C19$]". Storytelling toolkits for immersive data interfaces must therefore enable not only visual engagement, but also narrative clarity and pacing.

In summary, insights from our work help inform future research directions for immersive data storytelling, including the development of adaptive narrative logic, improved interaction affordances, and design methodologies for integrating contextual media into spatial storytelling experiences. 